\pdfoutput=1
\documentclass[a4paper]{article}
\usepackage{colm2024_conference}
\usepackage[most]{tcolorbox}
\usepackage{microtype}
\usepackage{hyperref}
\usepackage{todonotes}
\usepackage{url}
\usepackage{graphicx}
\usepackage{subcaption}
\usepackage{booktabs}
\usepackage{amssymb}
\usepackage{authblk}
\usepackage{hyperref}
\let\svthefootnote\thefootnote
\newcommand\freefootnote[1]{%
  \let\thefootnote\relax%
  \footnotetext{#1}%
  \let\thefootnote\svthefootnote%
}
\definecolor{darkblue}{rgb}{0, 0, 0.5}
\hypersetup{colorlinks=true, citecolor=darkblue, linkcolor=darkblue, urlcolor=darkblue}

\title{Generating Diverse Criteria On-the-Fly to Improve Pointwise LLM Rankers}

\author[1*]{\small Fang Guo}
\author[2*]{\small Wenyu Li }
\author[3]{\small Honglei Zhuang}
\author[1]{\small Yun Luo}
\author[1]{\small Yafu Li}
\author[4]{\small Qi Zhu}
\author[3]{\small Le Yan}
\author[1]{\small Yue Zhang}
\affil[1]{\footnotesize Department of Engineering, Westlake University}
\affil[2]{\footnotesize Department of Future Technology, South China University of Technology}
\affil[3]{\footnotesize Google Reserach}
\affil[4]{\footnotesize Department of Computer Science, University of Illinois at Urbana-Champaign}
\affil[1]{\textit {\{guofang,luoyun,liyafu,zhangyue\}@westlake.edu.cn}}
\affil[2]{\textit {\{wenyulilwy\}@gmail.com}}
\affil[3]{\textit {\{hlz,lyyanle\}@google.com}}
\affil[4]{\textit {\{qiz3\}@illinois.edu}}
\vspace{-3mm}

\begin{document}

\maketitle
\begin{abstract}
\freefootnote{* These two authors contribute equally.}
The most recent pointwise Large Language Model (LLM) rankers have achieved remarkable ranking results. However, these rankers are hindered by two major drawbacks: (1) they fail to follow a standardized comparison guidance during the ranking process, and (2) they struggle with comprehensive considerations when dealing with complicated passages. To address these shortcomings, we propose to build a ranker that generates ranking scores based on a set of criteria from various perspectives. These criteria are intended to direct each perspective in providing a distinct yet synergistic evaluation. Our research, which examines eight datasets from the BEIR benchmark demonstrates that incorporating this multi-perspective criteria ensemble approach markedly enhanced the performance of pointwise LLM rankers\footnote{Code, data, and prompts can be found at: \href{https://github.com/fangguo1/MCRanker}{https://github.com/fangguo1/MCRanker}}.
\end{abstract}

\section{Introduction} 

The integration of Large Language Models (LLMs) into various text rankers has resulted in notable advancements, outperforming traditional neural ranking approaches even in a zero-shot setup ~\citep{qin2023large,sun2023chatgpt,zhuang2023rankt5,pradeep2023rankvicuna,li2023parade}. Nevertheless, the iterative nature of querying LLMs, combined with their stochastic behavior, leads to inconsistent assessment criteria for zero-shot rankers, resulting in a lack of both \textit{consistency} and \textit{comprehensiveness} in the ranking process.

\begin{figure}
    \centering
    \includegraphics[scale=0.4]{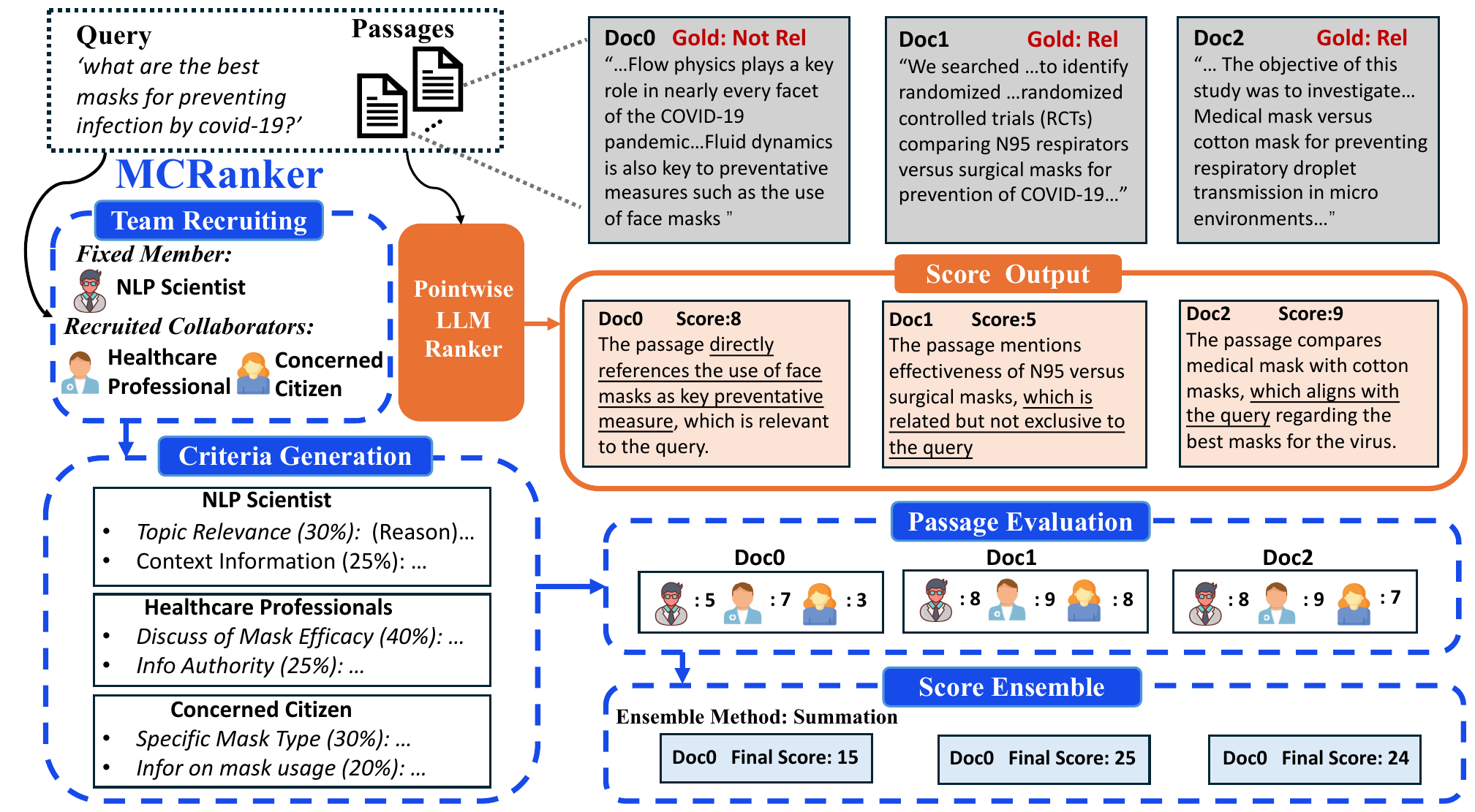}
    \caption{Pipeline of the proposed MCRanker in \textcolor{blue}{blue dashed line} and the example output of the Pointwise LLM-based Ranker in \textcolor{orange}{orange straight line}.}
    \label{fig1}
    \vspace{-2mm}
\end{figure}

In the realm of LLM rankers, a pointwise ranker, which evaluates each passage individually without comparing its position or relation to other passages, owns its advantage in scalability and interpretability \citep{liang2022holistic,zhuang2023beyond,sachan2022improving}. However, the pointwise ranker is not immune to the issues of inconsistent and biased assessments of passages. Figure \ref{fig1} illustrates such a concrete example in \textcolor{orange}{orange straight line}. The inconsistency happens when the contrasting scores are assigned to content with similar semantics. For instance, both Document 1 and Document 2 discuss the comparative effectiveness of various mask types in preventing COVID-19 transmission but are assigned markedly different scores by the pointwise ranker. 
Moreover, Document 0, which is tangentially related to the topic erroneously receives a high score. This error happens when the LLM ranker decides to adopt a biased assessment criterion that prioritizes keyword presence over the nuanced understanding of content semantics. In this paper, we investigate how to build a zero-shot pointwise ranker that is capable of generating both consistent and comprehensive assessments of passages. 

Recent studies on human evaluative practices ~\citep{BloombergBestPractise,gehrmann2023repairing} suggest that optimal annotation outcomes are achieved through the collaboration of annotators with diverse expertise, bolstered by a standardized annotation guideline. On one hand, optimal annotations necessitate a synergy between domain-specific experts and language experts to guarantee the integrity of the evaluation. For example, when annotating the TREC COVID dataset ~\citep{treccovid}, the annotation team included both medical students and professional indexers. On the other hand, the absence or ambiguity of annotation criteria could significantly diminish the quality and efficiency of the annotation process ~\citep{van2019best,gehrmann2023repairing}. Consequently, it is imperative to establish precise, well-articulated criteria before the commencement of the annotation process.


In line with these guidelines for humans, we introduce the MCRanker framework, which generates \textbf{M}ulti-perspective \textbf{C}riteria to improve pointwise LLM \textbf{Ranker}.
Figure \ref{fig1} illustrates the pipeline of MCRanker for query \textit{``What are the best masks for preventing infection by COVID-19?''}, highlighted in \textcolor{blue}{blue dashed line}. After reading the query, MCRanker emulates the domain expertise and text analytical capabilities of professional annotators through virtual team recruiting. This team comprises a fixed ``NLP scientist'' and recruits two collaborators: a ``health professional'' and a ``concerned citizen'', each contributing their unique perspective to the annotation process. Upon reviewing the query, each team member generates a set of weighted criteria reflective of their viewpoints. For example, the NLP scientist emphasizes text analysis criteria such as ``topic relevance'' and ``keywords match'', while the health professional focuses on the specificity of mask-related information. In the passage evaluation phase, each team member independently assigns a score to the passage based on his established criteria. Finally, individual assessments are ensembled through aggregation, yielding a score that reflects the passage's relevance to the given query.


In summary, the key contributions of this paper are as follows:
\begin{itemize}

  \item We introduce MCRanker, the first model to integrate a multi-perspective ideology into a ranking task. It simulates the professional human annotation process by building an annotation team that has both text analysis skills and diverse domain expertise.
  \vspace{-1pt}
  \item We investigate how criteria can be generated and utilized to influence a ranker. The in-depth analysis further shows the necessity of query-centric criteria to enhance ranking performance. 
  \vspace{-1pt}
  \item We evaluate MCRanker on 8 datasets from the BEIR benchmark. The results demonstrate that our proposed approach consistently delivers superior ranking results across various datasets.

\end{itemize}

\section{Related Work} 
\textbf{Text Ranking} Recent research has been investigating the application of large language models (LLMs) for zero-shot text ranking 
~\citep{nogueira2019multi,han2020learning,zhuang2021ensemble,nogueira2020document,zhuang2023rankt5,xian2023learning,liu2009learning,qin2020neural,zhang2023rankwithout}. Notably, pairwise ~\citep{qin2023large} and listwise ~\citep{sun2023chatgpt,ma2023zero,zhuang2023setwise,pradeep2023rankvicuna,pradeep2023rankzephyr} LLM rankers involve the simultaneous evaluation of two or more documents to generate a ranked list. However, these ranking strategies typically necessitate the establishment of a quality initial document order, which is often provided by a first-stage ranker, such as a pointwise ranker.

Pointwise rankers evaluate each query-passage pair separately, offering benefits in terms of scalability and interpretability. Several pointwise rankers derive the relevance score from the likelihood of the document's relevance to the query \citep{liang2022holistic} or the probability of generating the query from the document \citep{sachan2022improving}. \citeauthor{zhuang2023beyond} leverages fine-grained relevance labels within prompts to enable more nuanced differentiation among documents. Nevertheless, these models face consistency issues and cannot make comprehensive passage assessments. 

\textbf{Zero-shot LLM Assessors} Recent studies have explored the LLMs' capability as pseudo-assessors~ \citep{faggioli2023perspectives,faggioli2024determines,raina2024llmasajudge,wang2023doris,thomas2023large}. The primary objective of these LLM assessors is to assign a relevance label to each query-passage pair. Ideally, these relevance labels should be in line with human-generated ground-truth relevance labels. 
For example, \citeauthor{thomas2023large} designed prompts that incorporate several handcrafted aspects while \citeauthor{wang2023doris} have introduced a pipeline to let LLMs perform expert-level dataset annotation.

LLM assessors are built to establish evaluation datasets, marking a distinct departure from the goal of LLM rankers. LLM rankers primarily focus on ensuring the accuracy of the relative ordering among the top-ranked documents, without necessarily providing explicit relevance labels for each query-passage pair.

\textbf{Multi-Perspective Systems} The concept of ``Multi-Perspective Problem Solving'' has garnered enormous attention in recent research, particularly within the domains of ``Multi-Agent'' systems~\citep{deshpande2023toxicity,xu2023expertprompting,fu2023improving,li2023camel,wang2024rethinking,nair2023dera} and ``Mixture-of-Experts'' systems~\citep{li2022branchtrainmergemoe}, demonstrating its potential in resolving complex tasks. Notably, recent advancements in multi-agent systems have demonstrated the capacity of LLMs to assume specific identities, either through automatic assignment~\citep{wang2023unleashing,shao2024assisting} or manual selection~\citep{lan2023stance,zhao2024longagent, zhang2023exploring}. These identified LLM agents follow various collaboration mechanisms~\citep{chen2024reconcile} to accomplish given tasks. 

Inspired by this ``Multi-Perspective'' philosophy, we design a ``Team Recruiting'' step, which automatically generates multiple collaborators to work with a fixed NLP scientist. In this work, we did not study the collaboration mechanism within the team and leave it for future work. 
\section{Methods} 
In this section, we first revisit pointwise LLM rankers. Then we introduce our proposed MCRanker in detail. The pipeline of MCRanker is shown in Figure \ref{fig1}.

\subsection{Preliminary} 
We formally describe how a pointwise LLM ranker tackles a ranking problem. Given a specific query denoted as $q$, along with a set of passages to be evaluated as $P = (p_1, \ldots, p_m)$. The pointwise ranking function, represented as $f$, evaluates each pair consisting of the query $q$ and an individual passage $p_i$. It computes a score $s_i=f(q, p_i)$, signifying each passage's relevance to the query. After the pointwise ranker has computed relevance scores for all query-passage pairs $(q, p_i)$, it ranks all passages $P$ based on predicted scores $S = (s_1, \ldots, s_m)$ in descending order and outputs the ranked list as the final result. Notice that pointwise ranker can be subdivided into likelihood-based and text-based. Likelihood-based methods derive ranking scores from the generative likelihood of LLMs, whereas text-based methods derive their ranking scores directly from the textual outputs of the LLMs. In this study, we concentrate on the text-based pointwise ranker because 
token likelihood is in general unavailable in closed-source LLMs like GPT.
However, likelihood-based LLM rankers can be seamlessly plugged into our proposed framework.

\subsection{Method}
To make a pointwise ranker evaluating query-passage pairs in consistent and comprehensive criteria, we propose MCRanker which instructs LLM to perform the following procedures: (1) \textbf{Team Recruiting} that generates a virtual annotation team based on the query. This team includes a designated NLP scientist and a few collaborators with different domain expertise to provide various viewpoints for the upcoming annotation step; (2) \textbf{Criteria Generation} that prompts each team member to create detailed criteria for the upcoming evaluation step; (3) \textbf{Passage Evaluation} that lets each team member evaluate query-passage pairs following his criteria and (4) \textbf{Score Ensemble and Ranking} that ranks on the final scores, which are ensembled from team members' evaluation results. Figure \ref{fig1} shows a working example of MCRanker during inference.

\textbf{Team Recruiting} The first step is to establish a virtual team endowed with expert-level annotation capabilities, encompassing domain-specific expertise and text analysis proficiency. To acquire domain knowledge, we prompt a team recruiting LLM $\textit{M}_{Recruit}$ to let it decide each team member's identity. The prompt is designed to ask $\textit{M}_{Recruit}$ to guess who comes out of the given query. To ensure a comprehensive skill set across the team, the prompt fosters an interdisciplinary mix of talents. In alignment with our analysis of the human annotation process in the preceding ``Introduction'' section, our virtual annotation team also necessitates advanced text analysis skills. However, we empirically find that $\textit{M}_{Recruit}$ always fails to recruit a professional in text and language. Therefore, we designate an NLP scientist to join the team. The fully formed annotation team $A$ is made up of an NLP scientist $a_0$ and other collaborators $a_1,\ldots,a_n$. This recruiting process can be described as: 
\begin{equation}
    {a_1,\ldots,a_n}=\textit{M}_{Recruit}(x_r(q))
    \label{eq1}
\end{equation}
where $x_r$ is the prompt for team recruiting that takes the query as input. It can be found in Appendix \ref{Team Recruiting}.

\textbf{Criteria Generation} In the process of professional annotation, adherence to established scoring criteria is crucial for annotators to maintain uniformity in their evaluations. In light of this, our virtual annotation team mandates that each member formulate their own set of scoring criteria. These query-centric criteria are also expected to include a weighted distribution for each criterion, ensuring a systematic assessment. This step is formulated as:
\begin{equation}
    c_j=\textit{M}_{Criteria}(x_c(q,a_j))
    \label{eq2}
\end{equation}
where $0 \leqq j \leqq n$, $\textit{M}_{Criteria}$ is the LLM responsible for criteria generation. $x_c$ is the corresponding prompt for $\textit{M}_{Criteria}$  and can be found in Appendix \ref{NLP Scientist Criteria Generation} and \ref{Team Member Criteria Generation}.

\textbf{Passage Evaluation} The member of the annotation team then starts to evaluate the passages. For each query-passage pair $(q,p_i)$, the evaluation is fulfilled through prompting a passage evaluation LLM $\textit{M}_{Evaluation}$. It processes by letting each team member read the query, the passage as well as the query-centric criteria that he established in the previous step. Then each team member is expected to rate the relevance on a scale from 0 to $k$. This evaluation procedure can be described as:
\begin{equation}
    s_{ij}=\textit{M}_{Evaluation}(x_e(q,p_i,a_j,c_j))
    \label{eq3}
\end{equation}
where $x_e$ is the prompt for passage evaluation and can be found in Appendix \ref{Team Member Score}. 

\textbf{Score Ensemble} Once we get the passage evaluation result from each team member, we can ensemble their result to get a final score $s_i$. We consider three different ensemble methods: (1) Score Summation, (2) Rank Ensemble, and (3) LLM Assessor $\textit{M}_{assessor}$. 

For (1) Score Summation, we obtain the final score of a passage by simply adding up the scores from each team member.
For (2) Rank Ensemble, we first rank each team member $a_j$'s evaluation result to get a ranked list $r^j$. Then we calculate each passage's final score by summing up its mean reciprocal rank scores in each ranked list. For (3) LLM Assessor, we prompt $\textit{M}_{Assessor}$ by feeding in each team member's evaluation result on the passage and letting it give an overall score.

The three computing equations are shown below in order:
\begin{align}
    s_{\text{Sum}, i} &= \sum_{j=0}^{|A|} s_{ij}  & 
    s_{\text{RE}, i} &= \sum_{j=0}^{|A|} \frac{1}{r^j_{i}}   & 
    s_{\text{Assessor}, i} &= \textit{M}_{Assessor}(x_a(q,p_i,s_{i0},\ldots,s_{ij},r_{i0},\ldots,r_{ij}))
    \label{eq4}
\end{align}
where $x_a$ is the prompt for the final score assessment and is shown in Appendix \ref{LLM assessor}. The final output of MCRanker is a ranked list on $s_i$ in descending order.

\section{Experimental Setup}

\subsection{Dataset}
Same as \citep{zhuang2023beyond}, our experiments were conducted on 8 datasets from BEIR benchmark: Covid, Touche, DBPedia, SciFact, Signal, News, Robust04, and NFCorpus.


\subsection{Compared Methods}
We compared various zero-shot pointwise rankers as follows:
\begin{enumerate}
    

    \item \textbf{Query Generation (QG) \citep{sachan2022improving}}: This method rescores retrieved passages with a zero-shot question generation model. It uses a pre-trained language model to compute the probability of the input question conditioned on a retrieved passage. 

    \item \textbf{Binary Relevance Generation (RG-YN) \citep{liang2022holistic}}: It prompts the LLM to predict on a query-passage pair and uses the likelihood of ``Yes/No'' for ranking score calculation. 

    \item \textbf{Rating Scale 0-to-k Relevance Generation (RG-S) \citep{zhuang2023beyond}}: This method prompts the LLM to first output the relevance label for each query-passage pair, then calculates the expected relevance value using relevance value and corresponding marginal probability.

    \item \textbf{Rating Scale 0-to-k Directly Score (DIRECT(0, k))}: This method prompts the LLM to directly generate the relevance score for each query-passage pair. We adopt the prompt from \citeauthor{zhuang2023beyond} and k represents the rating scale. The prompt we use is shown in Appendix \ref{DIRECT(0, k)}.

\end{enumerate}
Notice that QG, RG-YN, and RG-S are all likelihood-based pointwise rankers, and DIRECT(0, k) is text-based.   

\subsection{Configurations}
We first used BM25 through pyserini\footnote{\href{https://github.com/castorini/pyserini}{https://github.com/castorini/pyserini}} to retrieve the top-100 documents from each query of every dataset, then ranked the retrieved documents with our MCRanker and the baseline methods. The ranking performance was measured by NDCG@10 \citep{jarvelin2002ndcg}.  

For MCRanker, we used ``GPT-4-1106-Preview'' as the base LLM model for $\textit{M}_{Recruit}$, $\textit{M}_{Criteria}$, $\textit{M}_{Evaluation}$ and $\textit{M}_{Assessor}$. The temperature is set to 0 and the rating scale k is set to 10. For DIRECT(0, k), "GPT-4-1106-Preview" served as the base LLM with the temperature similarly set to 0. For QG, RG-YN, and RG-S, we copied their performance on BEIR directly from the original papers. For $\textit{M}_{Evaluation}$, we prompt it to output a score without further explaining the reason (Appendix~\ref{Team Member Score}), this setup is in line with RG-S.

The ensemble mechanism used for MCRanker was set to ``Score Summation'' by default, namely the first equation in Equation \ref{eq4}. When needed, the method name was appended with the suffix ``-RE'' if calculated by Rank Ensemble, and ``-$\textit{M}_{Assessor}$'' by the LLM assessor.
\section{Results}


\subsection{Overall Performance}

\begin{table}[t]
\caption{
Overall ranking performances measured by average NDCG@10 on BEIR data sets. The best performances are bolded, and the second are underlined. NLP Sci. represents NLP Scientist, R.C. represents the Recruited Collaborator, and a suffix ${MC}$ represents our MCRanker with different amount of team members. * means likelihood-based models that utilize the token probability for relevance score calculation.
}
\centering
\small
\label{Table: Main results}
\setlength{\tabcolsep}{2pt}
\renewcommand{\arraystretch}{1.3}
\begin{tabular}{cccccccccc}
\toprule
\multicolumn{1}{c|}{\textbf{Method}}     &\textbf{Covid}      & \textbf{Touche}    & \textbf{News} & \textbf{Signal} & \textbf{DBPedia} & \textbf{SciFact} & \textbf{Robust04} & \textbf{NFCorpus} & \textbf{Avg}\\ \bottomrule
\multicolumn{1}{c|}{QG*}          & 73.57   & 24.08    & 41.56   & 28.72   & 37.73   & \underline{74.95}   & 46.51   & 36.73     & 45.48 \\
\multicolumn{1}{c|}{RG-YN*}       & 78.97   & 24.27    & 45.88  & \underline{31.96}   & 36.96   & 69.58  & 56.56   & 37.43     & 47.70   \\ 
\multicolumn{1}{c|}{RG-S*}  & 80.48   & 27.57    & 47.90   & \textbf{33.01}   & 41.90  & \textbf{75.21}  & 56.68   & 39.01     & 50.22 \\\bottomrule
\multicolumn{1}{c|}{DIRECT(0, 10)}     & 79.30   & 25.22    & 46.19   & 29.12   &  40.82  &  70.08  &  53.78  &   37.52   & 47.75 \\
\multicolumn{1}{c|}{DIRECT(0, 20)}       & 79.96   & 22.05  & 47.57  & 27.77     &   40.53  &  70.53   &  54.66   &  37.19   & 47.53  \\ \bottomrule
\multicolumn{1}{c|}{(NLP Sci.)$_{MC}$}     &  81.49   &   29.33 & 46.13    &   29.48   &  41.25  &  70.86  &  56.37   &    38.25   &   49.14 \\ 
\multicolumn{1}{c|}{(1 R.C.)$_{MC}$}     &   82.43  &   30.60   &  48.52  &  26.58  &  	41.11  &  71.35  &  55.78   &    37.93   &   49.28 \\ 
\multicolumn{1}{c|}{(2 R.C.)$_{MC}$}     &  83.53   &  31.42   &  \textbf{50.90}   &  26.85  &  42.33  &  71.86  &  56.84   &   38.36    &  50.26  \\ 
\multicolumn{1}{c|}{(NLP Sci. + 1 R.C.)$_{MC}$}     &  \underline{84.16}  &  \textbf{32.91}    &   49.54  &   29.94  &  \underline{43.85}  &  73.33  &  \underline{57.12}   &   \underline{39.12}    &  \underline{51.24}  \\ 
\multicolumn{1}{c|}{(NLP Sci. + 2 R.C.)$_{MC}$}     & \textbf{84.23}   & \underline{32.48}    & \underline{50.32}   & 29.73   & \textbf{44.67}   &  73.14  &  \textbf{57.23}    &   \textbf{39.58}    &  \textbf{51.42}  \\ 
\bottomrule
\end{tabular}
\end{table}

Table \ref{Table: Main results} summarizes the overall performance on eight datasets from BEIR. Our method got the best performance and outperformed the direct GPT4-prompting ranker DIRECT(0, 10) in average performance by nearly 8$\%$ in NDCG@10.

Upon analyzing the table, several observations can be inferred: (1)The method introduced in this study, MCRanker, which incorporates a virtual annotation team comprising an NLP scientist and one or two additional collaborators, consistently outshines the baseline across all datasets. When compared with the direct score baseline DIRECT, MCRanker displays a remarkable improvement, registering an average increase in NDCG@10 by a magnitude of 3.67. This enhancement underscores the efficacy of MCRanker in augmenting the LLM's capability for more accurate relevance predictions. (2) The variant of MCRanker that utilizes additional two collaborators besides an NLP scientist (NLP Sci. + 2 R.C.)$_{MC}$ shows an increase in performance over the single-collaborator one (NLP Sci. + 1 R.C.)$_{MC}$ in five datasets, although this improvement is not statistically substantial. This may be attributed to the overlapping expertise between the NLP scientist and the other collaborators. A more granular analysis of the multi-perspective annotation will be provided in Section \ref{multi-perspective ablation}. (3) 
A comparative examination of the results from DIRECT with rating intervals of $(0,10)$ and $(0,20)$ reveals that simply expanding the scale does not enhance the LLM's ability to discriminate between relevant and non-relevant passages. This finding further demonstrates the performance gain observed in MCRanker can be ascribed to the synergistic evaluation from each perspective. (4) MCRanker achieves higher scores than likelihood-based ranker RG-S in six out of the eight datasets, except the Signal and SciFact datasets. We hypothesize that the brevity of the Twitter posts in Signal and the ambiguous query passage relevance in SciFact may result in a confounding effect on the multi-perspective criteria evaluation, thereby impairing MCRanker's performance. Also, we hypothesize that if we could get the token probability from the LLM, the performance of MCRanker can be further enhanced.

\subsection{Ablation Study}
To investigate the impact of each module in our design, we conducted ablation studies to assess the performance of our framework when each module was removed or switched to a different configuration. In the following, we provide a detailed description of the results.

\begin{figure}
    \centering
    \includegraphics[width=1\linewidth]{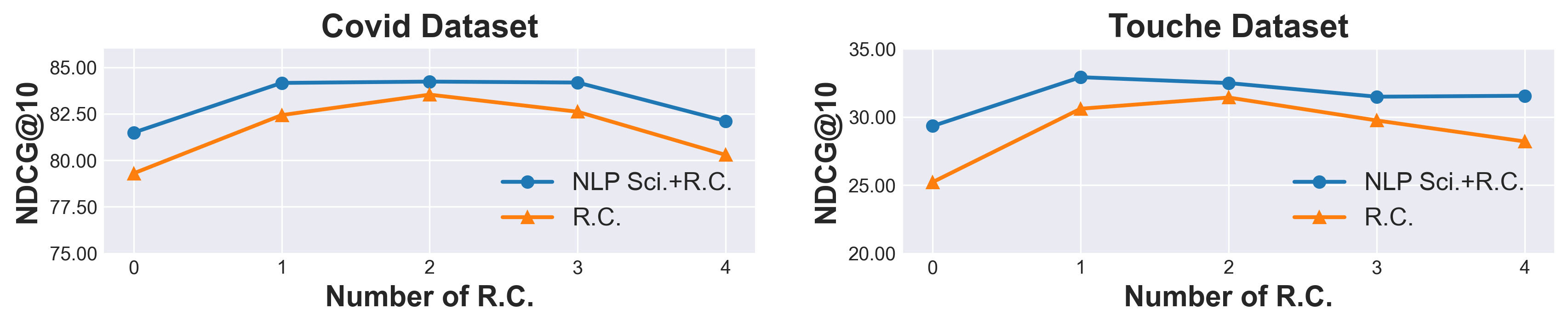}
    \caption{Research on different number of Team Member}
    \label{fig:team member number}
    \vspace{-3mm}
\end{figure}
\subsubsection{Study on multi-perspective annotation}
\label{multi-perspective ablation}
The virtual annotation team in the MCRanker framework includes a designated NLP scientist and a few recruited collaborators. Our experiments involved different combinations of team members to evaluate the efficacy of our approach. 


The results indicate that the inclusion of any team member like only one NLP scientist variant (NLP Sci.)$_{MC}$ or only one recruited collaborator (1 R.C.)$_{MC}$ contributes to a performance improvement when compared with the direct prompt baseline DIRECT(0,k). This suggests that anchoring the annotation criteria to a specific perspective yields more consistent and accurate predictions. Moreover, two recruited collaborators (2 R.C.)$_{MC}$ consistently overperform (1 R.C.)$_{MC}$ across all eight datasets. This finding supports the notion that an extra perspective can provide a more holistic evaluation, leading to performance gains. While the NLP scientist alone (NLP Sci.)$_{MC}$ achieves performance comparable to the single recruited collaborator variant (1 R.C.)$_{MC}$ on average, a notable performance gain is observed when the (NLP Sci. + 1 R.C.)$_{MC}$ model is compared to the (2 R.C.)$_{MC}$ variant. The former shows a greater synergistic effect. We believe it is due to the language model's preference to select annotators with expertise closely aligned with the query, even when prompted to consider interdisciplinary backgrounds. An NLP scientist undoubtedly contributes essential and valuable text analysis skills to the annotation team. 

A further experiment was conducted to examine the impact of varying the number of team members. As shown in Figure~\ref{fig:team member number}, on both datasets, as the number of R.C. increases, the optimal performance is reached when this number is 2, then the performance starts to drop. This might be due to redundancy in expertise among an increased number of collaborators. We leave the investigation of the underlying cause for future study.


\begin{table}[t]
    \caption{Ablation study on criteria utility measured by average NDCG@10.}
    \vspace{-3pt}
    \label{Table: Ablation study on criteria utility}
    \centering
    \small
    \setlength{\tabcolsep}{5pt}
    \renewcommand{\arraystretch}{1.2}
    \begin{tabular}{ccc}
      \toprule
      \textbf{Method} & \textbf{Covid} & \textbf{Touche} \\
      \midrule
      DIRECT(0, 10) & 79.30 & 25.22 \\
      DIRECT(0, 10)+criteria & 79.86 & 30.46 \\
      MCRanker without criteria & 79.97 & 26.22 \\
      MCRanker & \textbf{84.23} & \textbf{32.48} \\
      \bottomrule
    \end{tabular}
\end{table}
\begin{table}[t]
  \centering
  \begin{minipage}{0.47\textwidth}
    \caption{Ablation study on broad creteria measured by average NDCG@10. ``DBC'' represent ``Dataset-Based-Criteria''.}
    \vspace{-3pt}
    \label{Table: Ablation study on criteria generation}
    \centering
    \small
    \setlength{\tabcolsep}{5pt}
    \renewcommand{\arraystretch}{1.2}
    \begin{tabular}{ccc}
      \toprule
      \textbf{Method} & \textbf{Covid} & \textbf{Touche} \\
      \midrule
    DIRECT(0, 10) & 79.30 & 25.22 \\
      MCRanker-DBC & 82.28 & 27.78 \\
      MCRanker & \textbf{84.23} & \textbf{32.48} \\
      \bottomrule
    \end{tabular}
  \end{minipage}
  \hfill
  \begin{minipage}{0.47\textwidth}
    \caption{Ablation study on ensemble mechanism measured by average NDCG@10.}
    \vspace{-3pt}
    \label{Table: Ablation study on ensemble mechanism}
    \centering
    \small
    \setlength{\tabcolsep}{5pt}
    \renewcommand{\arraystretch}{1.2}
    \begin{tabular}{ccc}
      \toprule
      \textbf{Method} & \textbf{Covid} & \textbf{Touche} \\
      \midrule
      MCRanker-RE & 82.55  & 30.73  \\
      MCRanker-$\textit{M}_{Assessor}$ &  83.58 &  31.35 \\
      MCRanker & \textbf{84.23} & \textbf{32.48}   \\
      \bottomrule
    \end{tabular}
  \end{minipage}
\end{table}
\subsubsection{Study on criteria generation and utility}



To assess the impact of criteria on model performance, we employed a systematic ablation study. This involves integrating criteria developed by the virtual annotation team into DIRECT(0,10), removing all criteria from MCRanker, and transitioning from query-based to dataset-based criteria. 

Our findings, detailed in Table \ref{Table: Ablation study on criteria utility}, reveal that incorporating the same criteria used by MCRanker into DIRECT(0,10) yields improvements on all three datasets, with an especially significant enhancement of approximately 5.24 of NDCG@10 on Touche. Conversely, stripping MCRanker of its criteria results in a marked performance decline, with decreases of 4.26 and 6.26 of NDCG@10 respectively. These results underscore the critical role criteria play in not only guiding rankers in score generation but also ensuring consistency across evaluations. Absent these criteria, scores tend to decrease, reflecting a lack of standardization that can easily lead LLM to adopt inconsistent scoring guidelines.

Additionally, our analysis also investigates the efficacy of broad criteria. We keep all other steps the same in MCRanker except shifting from query-based to dataset-based criteria in $\textit{M}_{Criteria}$.
As Table \ref{Table: Ablation study on criteria generation} demonstrates, MCRanker-DBC (``DBC'' represents ``Dataset-Based-Criteria'') which adopts a dataset-level criteria has a performance decrease in approximately 1.95 of NDCG@10 for Covid and 4.70 for Touche. This decline illustrates the inherent complexity within the benchmark, where queries in a single dataset also diverge in semantics. It also suggests that a one-size-fits-all approach to criteria generation may not achieve optimal prediction performance and query-centric criteria generation step is necessary. Despite this decrease, MCRanker-DBC still records a notable performance improvement in comparison to DIRECT(0,10), demonstrating the importance of establishing criteria before scoring. This exploration can shed light on how to balance budget, specificity, and generality in criteria generation.

\subsection{Further Analysis}
In this section, we analyze the impact of different ensemble mechanisms, model performance with different rating scales, and the model's generalizability to different LLMs. Due to the space limit, we put two concrete examples of MCRanker's workflow in the Appendix Figure ~\ref{case1} and Figure ~\ref{case2}, and a visualization of the identities of the generated collaborators in the Appendix Figure ~\ref{fig:wordclouds}.


\subsubsection{Study on ensemble mechanism}
We examine the effect of various ensemble strategies on the performance of MCRanker.The results, detailed in Table \ref{Table: Ablation study on ensemble mechanism}, indicate that the straightforward "Score Summation" approach surpasses alternative methods. The limited effectiveness of the "Ranking Ensemble" method might be attributed to the homogeneity in scale of the scores derived from different team members. The LLM assessor, on the other hand, can be easily misled by certain members' scores, which requests for more delicate prompt engineering. Thus, we choose `` Score Summation'' as the principal ensemble method for its simplicity and robust performance.

\begin{figure}
    \centering
    \includegraphics[width=1\linewidth]{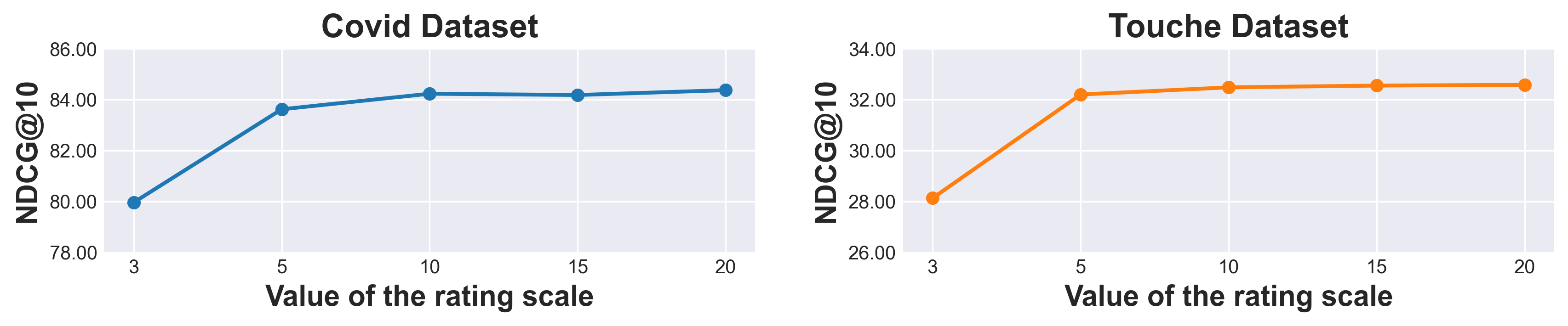}
    \caption{MCRanker with different values of the rating scale k}
    \label{fig:different_number_of_k}
    
\end{figure}

\subsubsection{Study on the variance of rating scale }
We plot how the performance changes about the rating scale for our proposed MCRanker in Figure \ref{fig:different_number_of_k}. On Covid and Touche datasets, when the rating scale k increases from 3 to 5, we can observe an apparent improvement. Then the performance continues to increase as k increases from 5 to 10 and becomes even as k reaches 15 and 20. These observations suggest that our methodology exhibits robustness to variations in different rating scales k when k is chosen within a reasonably large range.

\begin{figure}
    \centering
    \includegraphics[width=1\linewidth]{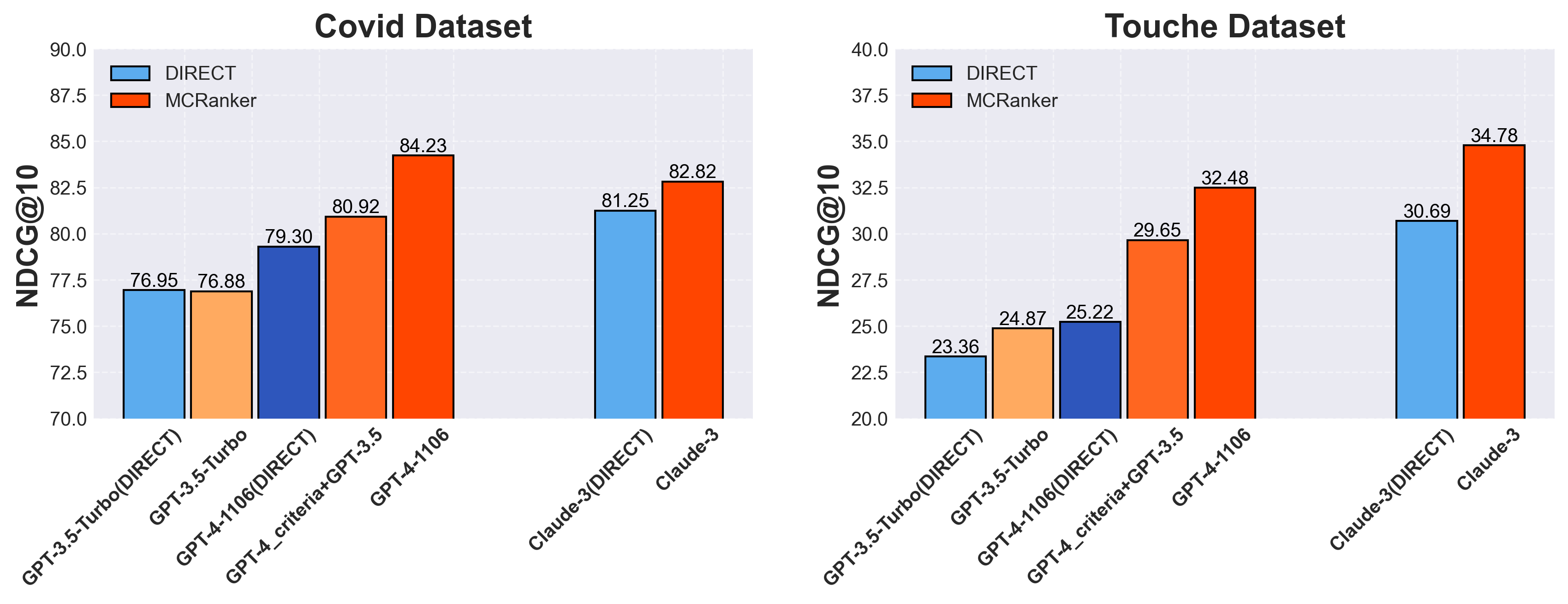}
    \caption{Comparing performance of different base models}
    \label{fig:different_model_comparel}
    \vspace{-3mm}
\end{figure}
\subsubsection{Generalizability to different LLMs}
To explore the generalizability of our proposed methods in different LLMs, we conducted experiments in which we replaced the base LLM with "GPT-3.5-Turbo" and "Claude-3-Sonnet". The result is presented in Table \ref{fig:different_model_comparel}. When using "GPT-3.5-Turbo" for all three modules: $\textit{M}_{Recruit}$, $\textit{M}_{Criteria}$, and $\textit{M}_{Evaluation}$, the performance decreases 7.35 in NDCG@10 on the Covid dataset and 7.61 on the Touche dataset. This level of performance is akin to using "GPT-3.5-Turbo" in a direct prompting baseline DIRECT(0,10). However, maintaining the "GPT-4-1106-Preview" model for $\textit{M}_{Recruit}$ and $\textit{M}_{Criteria}$, while only employing "GPT-3.5-Turbo" for $\textit{M}_{Evaluation}$, results in a substantial performance increase. This finding demonstrates that quality criteria can effectively guide a less advanced $\textit{M}_{Evaluation}$ to yield more accurate relevance predictions. Upon comparison between the "Claude-3-Sonnet" variant of MCRanker and the corresponding DIRECT(0,k) baseline, it is evident to see the performance increase. It underscores the robust generalizability of our proposed multi-perspective criteria ensemble methodology.

\section{Conclusion}
In this work, we explore the use of a multi-perspective criteria ensemble method to overcome the inconsistent and biased prediction from pointwise LLM rankers. Our experiments on BEIR benchmarks demonstrate our proposed method can consistently improve the ranking performance. This work is among the first to apply the concept of ``multi-perspective problem solving'' to a ranking task. Furthermore, our in-depth analysis reveals that quality criteria can robustly and significantly improve the performance of the pointwise LLM ranker, even when built upon a less powerful base model.

For future work, we will investigate how to extend our discovery to other ranking frameworks and further explore the collaboration mechanism behind the virtual annotators.

\newpage
\bibliography{colm2024_conference}
\bibliographystyle{colm2024_conference}

\appendix
\newpage
\section{Concrete Example of MCRanker}

\begin{figure}[h]
    \centering
    \includegraphics[scale=0.65]{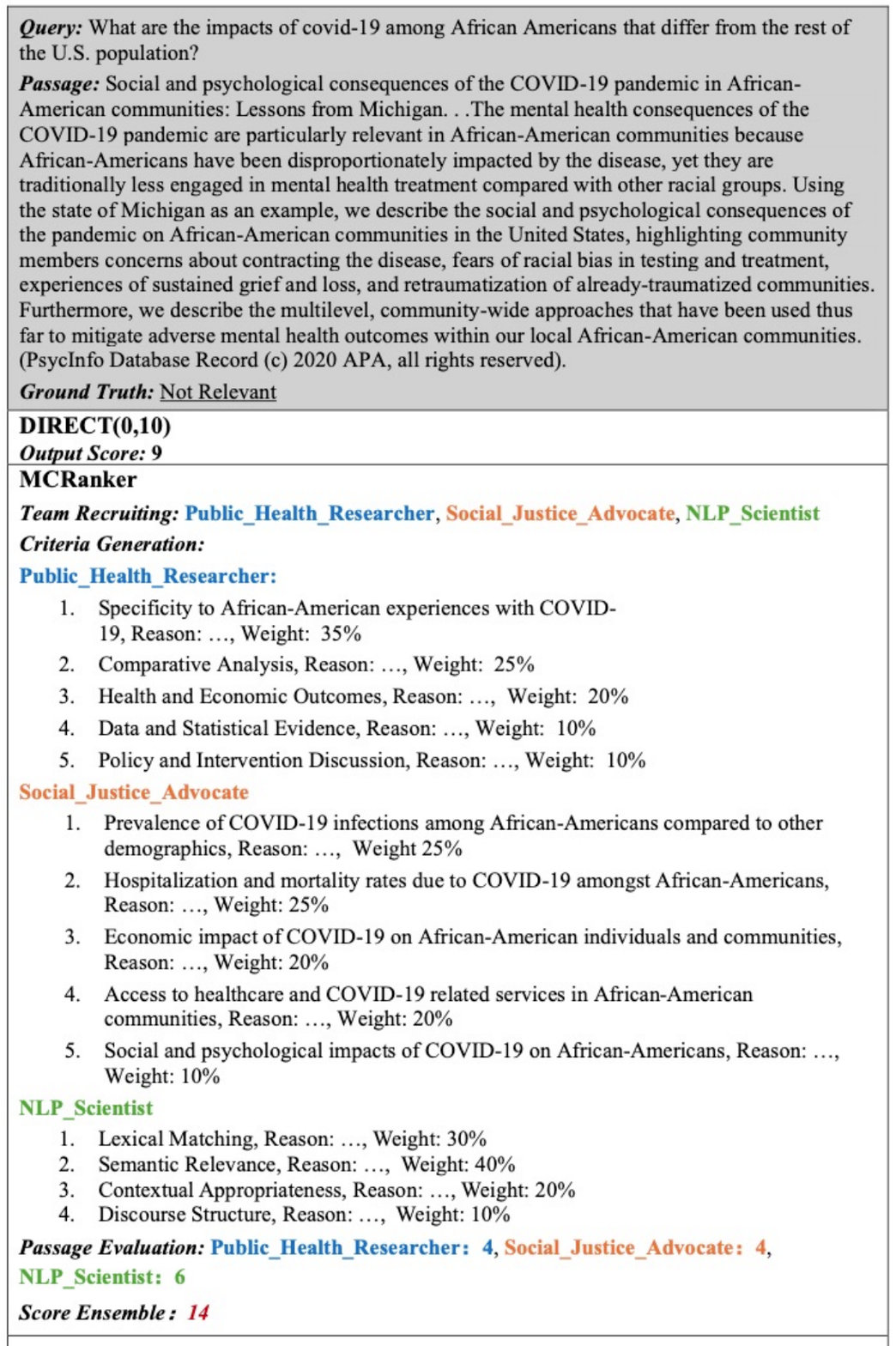}
    \caption{MCRanker spots a subtle semantic difference between the query and an irrelevant passage.}
    \label{case1}
\end{figure}
\begin{figure}[t]
    \centering
    \includegraphics[scale=0.65]{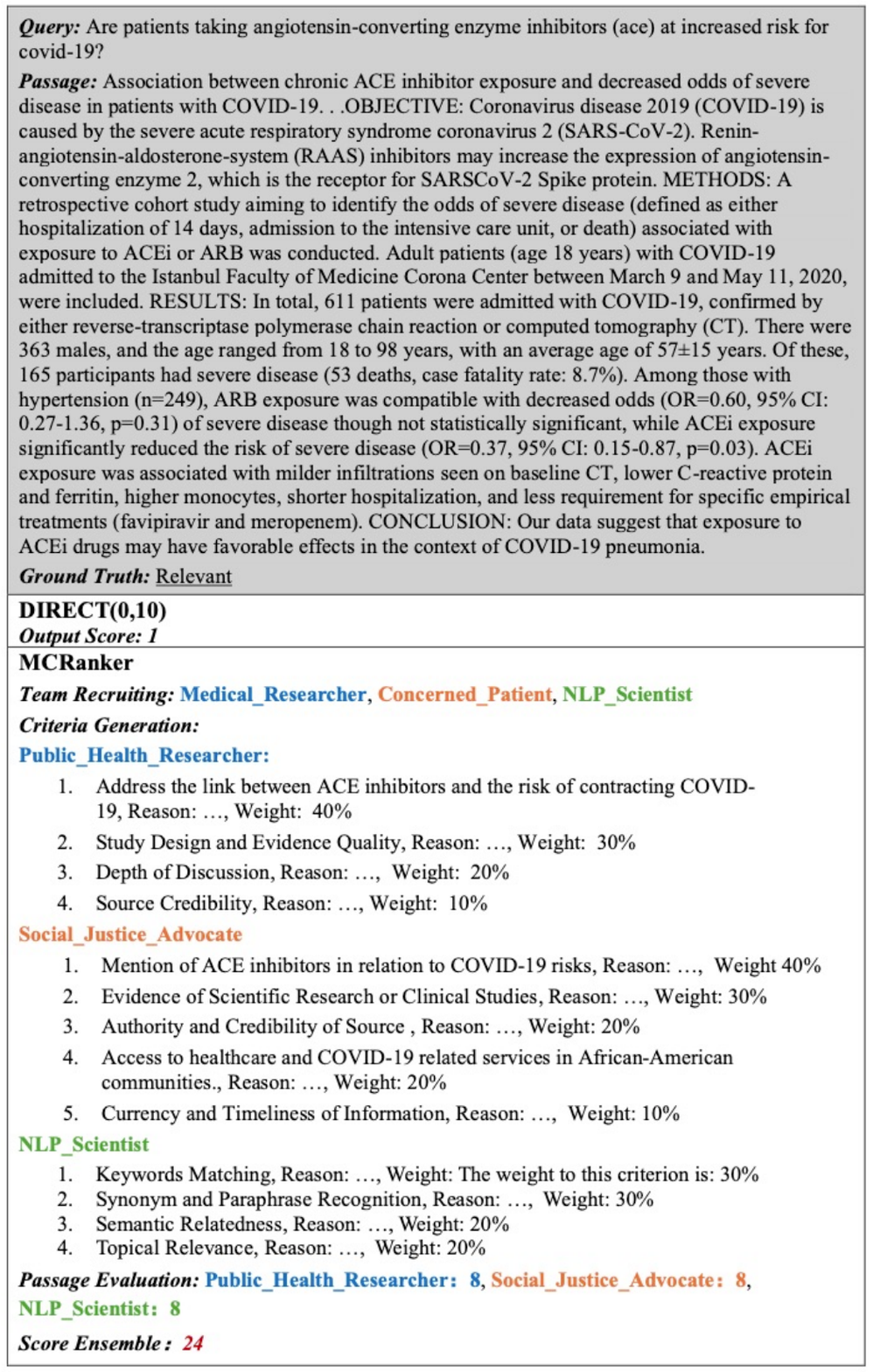}
    \caption{MCRanker's virtual annotation team conducts a more nuanced analysis of a relevant passage}
    \label{case2}
\end{figure}
We present two specific query-passage pairs from the COVID dataset to showcase the functionality of MCRanker's counterparts and highlight how they yield a score that diverges from that of the DIRECT baseline. The first case is shown in Figure~\ref{case1}. The query is about the comparison between the African-American community and the rest of the U.S. population during COVID-19 and an irrelevant passage is mainly about the consequences of COVID-19 in the African-American community. This subtle difference is spotted by MCRanker when most team members' criteria emphasize the ``comparison'' -related criterion. The second case, illustrated in Figure~\ref{case2}, involves a scenario where the DIRECT baseline incorrectly assigns a false negative prediction because the passage does not provide a direct response to the query. In contrast, MCRanker's virtual annotation team conducts a more nuanced analysis of the passage, leading to a high relevance score output.

\newpage
\section{More Comparison Results}
We also include a more thorough comparison with other methods including:
\begin{enumerate}

    \item \textbf{BM25}: The base retriever performance. 

    \item \textbf{monoT5}~\citep{nogueira2020monoT5}: A model based on T5 XL architecture and fine-tuned on the MS MARCO dataset. We test the model directly without further fine-tuning. 

    \item \textbf{RankT5}~\citep{zhuang2023rankt5}: This method uses an encoder-only model and employs a listwise softmax cross-entropy loss for ranking. It is fine-tuned on the MS MARCO dataset. We test the model directly without further fine-tuning. 

    \item \textbf{RankGPT}~\citep{sun2023chatgpt}: A zero-shot listwise LLM ranker that processes a query and a set of documents, producing a ranking of the documents based on their relevance. This method incorporates a sliding window approach. Besides the model that uses ``GPT-3.5-Turbo'' as the base LLM, we further include a ``GPT-4'' variant of it that does a second-stage ranking based on the ``GPT-3.5-Turbo'' version's ranking result.

\end{enumerate}

\begin{table}[h]
\caption{
Overall ranking performances measured by NDCG@10 on BEIR.\\
}
\centering
\scriptsize
\bfseries
\label{Table: More Comparison results}
\setlength{\tabcolsep}{2pt}
\renewcommand{\arraystretch}{1.3}
\begin{tabular}{cc|ccccccccc}
\toprule
\multicolumn{1}{c}{\textbf{Method}} & \multicolumn{1}{c|}{\textbf{Model}} & \textbf{Covid} & \textbf{Touche} & \textbf{News} & \textbf{Signal} & \textbf{DBPedia} & \textbf{SciFact} & \textbf{Robust04} & \textbf{NFCorpus} & \textbf{Avg}\\ \bottomrule
\multicolumn{1}{c}{BM25}           & N/A & 59.47   & 44.22    & 39.52   & 33.05   & 31.80   & 67.89   & 40.70   & 30.75     & 43.42 \\
\bottomrule
\multicolumn{1}{c}{QG}             & FLAN PaLM2 S & 73.57   & 24.08    & 41.56   & 28.72   & 37.73   & 74.95   & 46.51   & 36.73     & 45.48 \\
\multicolumn{1}{c}{RG-YN}          & FLAN PaLM2 S & 78.97   & 24.27    & 45.88   & 31.96   & 36.96   & 69.58   & 56.56   & 37.43     & 47.70   \\ \bottomrule
\multicolumn{1}{c}{RG-S(0,2)}           & FLAN PaLM2 S & 77.60   & 26.95   & 46.77   & 30.34   & 37.09   & 69.21   & 55.57   & 37.87     & 47.68 \\
\multicolumn{1}{c}{RG-S(0,4)}           & FLAN PaLM2 S & 80.48   & 27.57    & 47.90   & 33.01   & 41.90   & 75.21  & 56.68   & 39.01     & 50.22 \\\bottomrule
\multicolumn{1}{c}{DIRECT(0, 10)}   & GPT-4-1106-Preview & 79.30   & 25.22    & 46.19   & 29.12   &  40.82   &  70.08  &  53.78  &   37.52   & 47.75 \\
\multicolumn{1}{c}{DIRECT(0, 20)}   & GPT-4-1106-Preview & 79.96   & 22.05    & 47.57   & 27.77   &   40.53   &  70.53   &  54.66   &  37.19   & 47.53  \\ \bottomrule
\multicolumn{1}{c}{(NLP Sci.)$_{MC}$} & GPT-4-1106-Preview &  81.49   &   29.33 & 46.13    &   29.48   &  41.25   &  70.86   &  56.37   &    38.25   &   49.14 \\ 
\multicolumn{1}{c}{(1 R.C.)$_{MC}$}   & GPT-4-1106-Preview &   82.43  &   30.60   &  48.52   &  26.58   &  	41.11   &  71.35   &  55.78   &    37.93   &   49.28 \\ 
\multicolumn{1}{c}{(2 R.C.)$_{MC}$}   & GPT-4-1106-Preview &  83.53   &  31.42   &  50.90   &  26.85   &  42.33   &  71.86   &  56.84   &   38.36    &  50.26  \\ 
\multicolumn{1}{c}{(NLP Sci. + 1 R.C.)$_{MC}$} & GPT-4-1106-Preview &  84.16  &  32.91    &   49.54   &   29.94   &  43.85  &  73.33   &  57.12   &   39.12    &  51.24  \\ 
\multicolumn{1}{c}{(NLP Sci. + 2 R.C.)$_{MC}$} & GPT-4-1106-Preview & 84.23   & 32.48    & 50.32   & 29.73   & 44.67   &  73.14   &  57.23    &   39.58    &  51.42  \\ 
\bottomrule
\multicolumn{1}{c}{monoT5}    & Fine-tuned T5 XL & 80.71   & 32.41    & 48.49   & 32.55   & 44.45   & 76.57  & 56.71   & 38.97     & 51.36 \\
\multicolumn{1}{c}{RankT5}    & Fine-tuned T5 XL & 82.00   & 37.62    & 48.15   & 31.80   & 44.19   & 76.86   & 52.76   & 38.60     & 51.50 \\\bottomrule
\multicolumn{1}{c}{RankGPT}    &- GPT-3.5-Turbo & 76.67   & 36.18    & 48.85   & 32.12   & 44.47   & 70.43   & 50.62   & 35.62     & 49.37 \\
\multicolumn{1}{c}{}    &- GPT-4 & 85.51   & 38.47    & 52.89   & 34.40   & 47.12   & 74.95   & 57.55   & 38.47     & 53.68 \\
\bottomrule

\end{tabular}
\end{table}

From Table~\ref{Table: More Comparison results}, we can observe that MCRanker achieves compatible performance with fine-tuned rankers like monoT5 and RankT5. Also, it is encouraging to see that MCRanker outperforms listwise RankGPT with a ``GPT-3.5-Turbo'' base model and substantially shrinks the gap between the ``GPT-4'' version RankGPT.

\newpage
\section{Team Member Identity Visualization}
\begin{figure}[h]
    \centering
    \begin{subfigure}[b]{0.48\textwidth}
        \centering
        \includegraphics[scale=0.25]{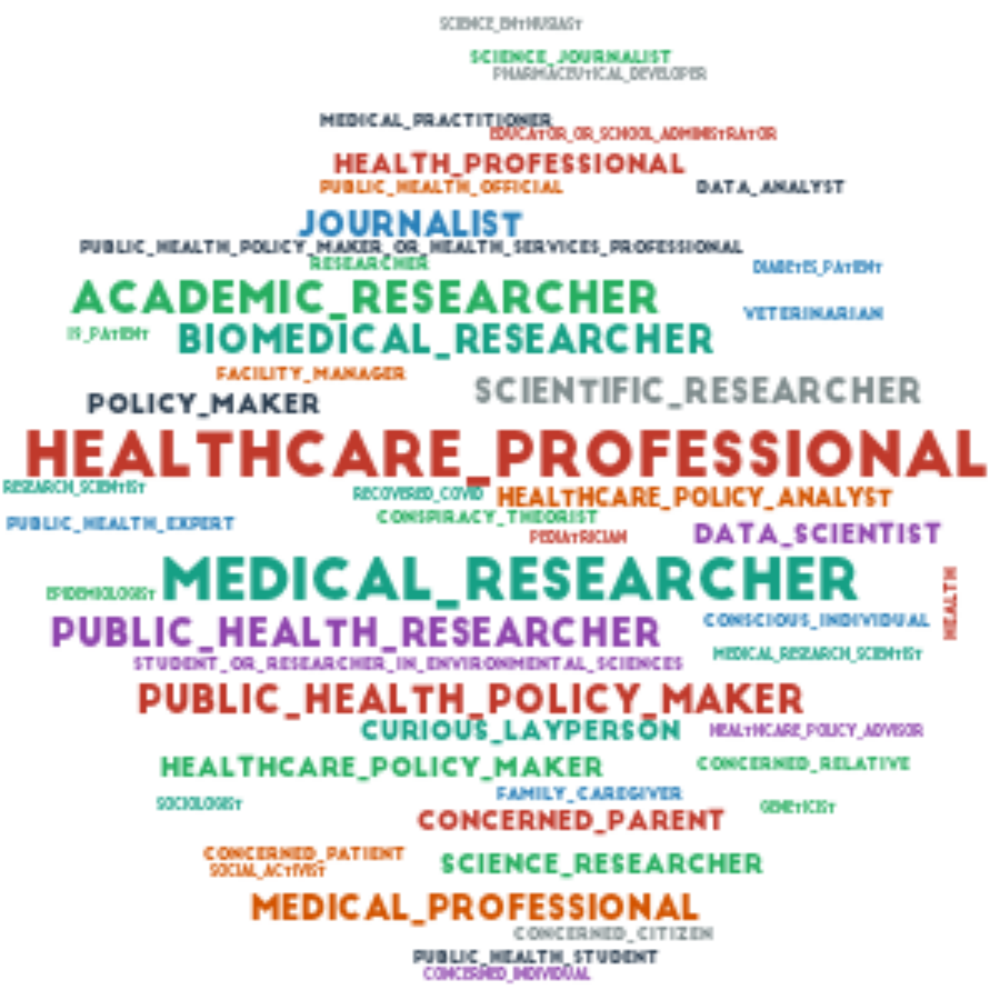}
        \caption{Identities for TREC-COVID}
        \label{fig:trec-covid}
    \end{subfigure}
    \hfill
    \begin{subfigure}[b]{0.48\textwidth}
        \centering
        \includegraphics[scale=0.25]{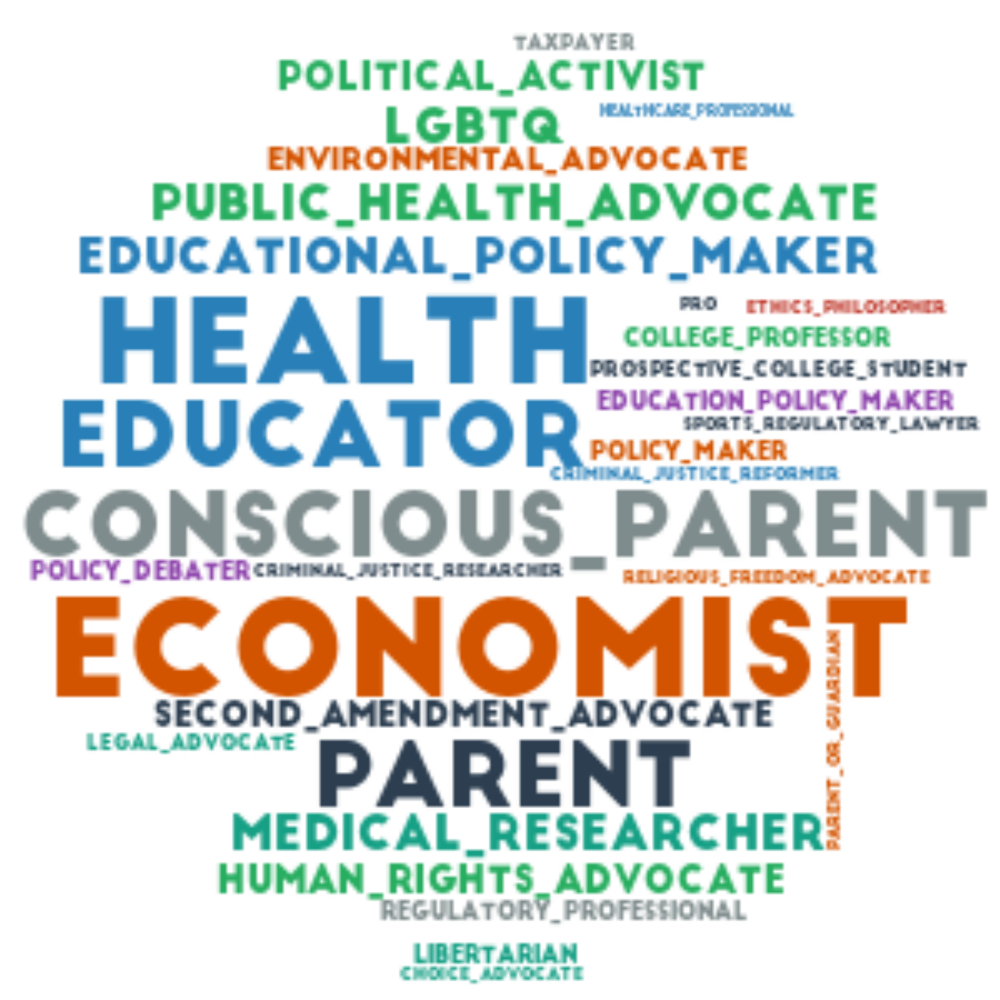}
        \caption{Identities for Webis-Touche2020}
        \label{fig:webis-touche2020}
    \end{subfigure}
    \vskip\baselineskip
    \begin{subfigure}[b]{0.48\textwidth}
        \centering
        \includegraphics[scale=0.25]{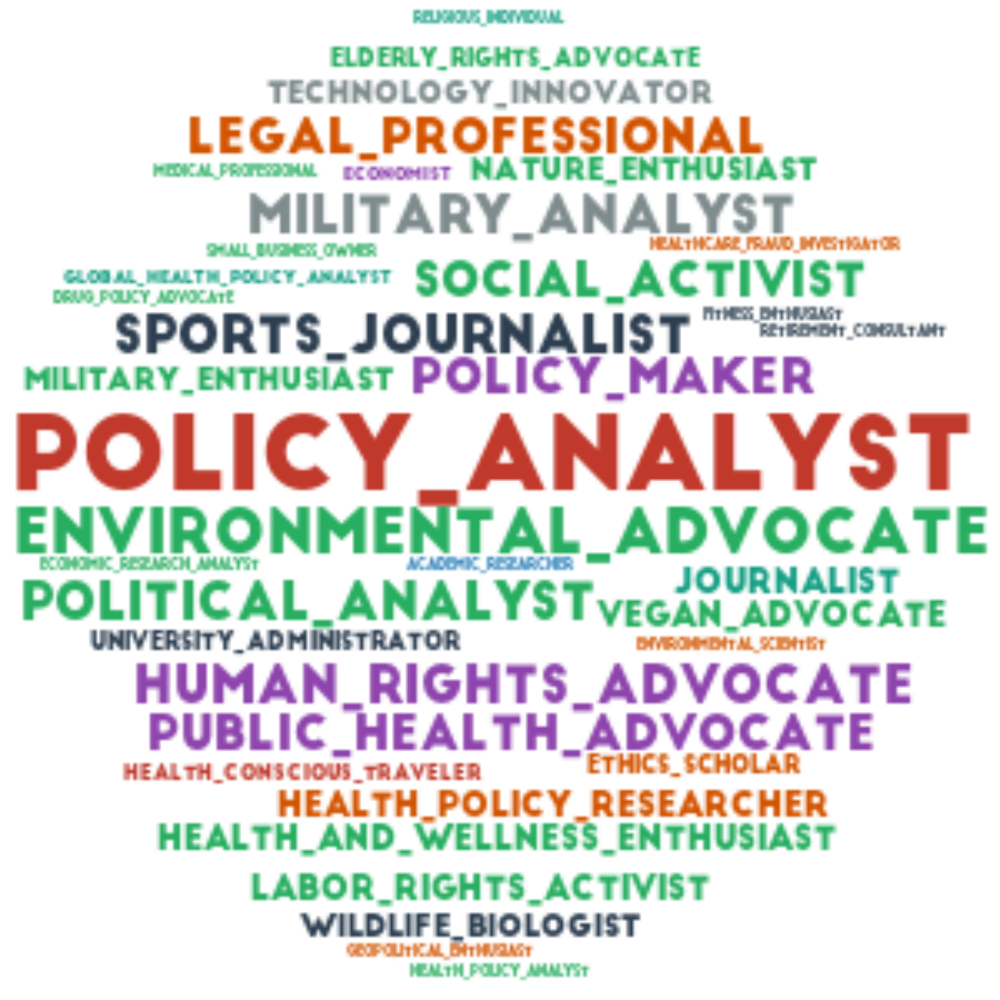}
        \caption{Identities for News}
        \label{fig:news}
    \end{subfigure}
    \quad
    \begin{subfigure}[b]{0.48\textwidth}
        \centering
        \includegraphics[scale=0.25]{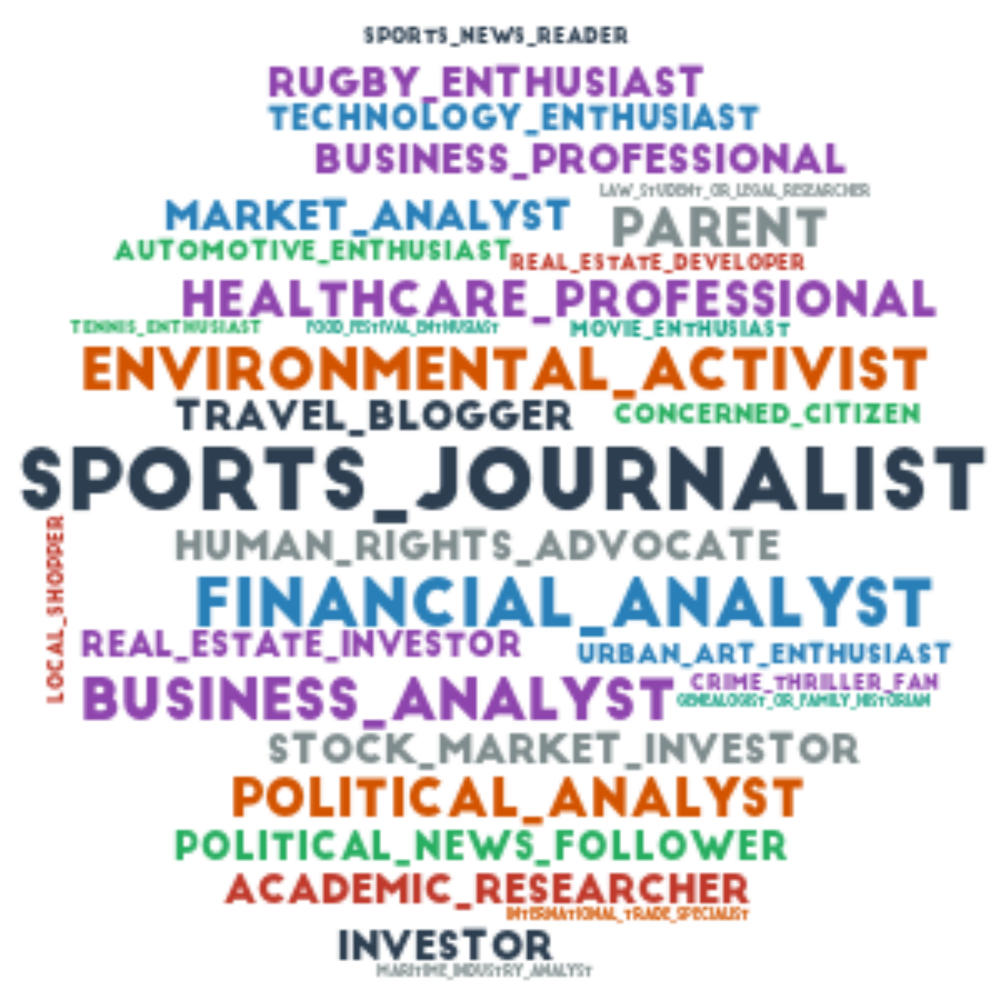}
        \caption{Identities for Signal}
        \label{fig:signal}
    \end{subfigure}
    \caption{Visualization of the identities of Recruited Collaborator (R.C.) on four datasets. The identities show a high correlation with the nature of the datasets.}
    \label{fig:wordclouds}
\end{figure}

4\section{Prompts}
\subsection{Team Recruiting}
\label{Team Recruiting}
\begin{tcolorbox}[breakable]
\bfseries
\small
The task is when given a query and a passage example, to try to guess the most probable user identities. You can name up to {number} identities. Try to make these identities very different from each other. Note that the passage example is just to show you the passage style and the user could possibly like/dislike it. So do not let the stance in the passage influence your guess.
\\\\\\
Given query:$<<<$\{query\}$>>>$
\\\\\\
Given passage:$<<<$\{passage\}$>>>$
\\\\\\
Please give your output in JSON format with keys are \textbackslash
``Identities\textbackslash
'' and \textbackslash
``Reason\textbackslash
'' .
\\\\\\
Under the content of \textbackslash
``Identities\textbackslash
'', please output the identity, your identity should be correct, clear,and easy to understand 
\\\\\\
Under the content of \textbackslash
``Reason\textbackslash
'', explain why you output these identities.
\\\\\\
Provide a clear and concise response, just give your answer in JSON format as I request, and don't say any other words.
\end{tcolorbox}


\subsection{NLP Scientist Criteria Generation}
\label{NLP Scientist Criteria Generation}
\begin{tcolorbox}[breakable]
\bfseries
\small
As an NLP Scientist, you are good at judging the linguistic relevance of a passage in relation to the given query, listing the criteria for judging linguistic relevance between the query and a passage, and assigning weights to each criterion. 
\\\\\\
Given query:$<<<$\{query\}$>>>$
\\\\\\
Please give your output in JSON format with keys are \textbackslash
``Criteria\textbackslash
'' and \textbackslash
``Reason\textbackslash
'' .
\\\\\\
Under the content of \textbackslash
``Criteria\textbackslash
'', please output the criteria and some explanation to it, your criteria should be correct, clear, executable, and easy to understand, the criteria should come from your knowledge as the the NLP Scientist. The weight to every criterion should be added after each criterion with 'The weight to this criterion is:'.
\\\\\\
Under the content of \textbackslash
``Reason\textbackslash
'', explain why you output these criteria.
\\\\\\
Focus solely on your NLP Scientist expertise and avoid deductions outside of your expert criteria. 
\\\\\\
Provide a clear and concise response, just give your answer in JSON format as I request, and don't say any other words.
\end{tcolorbox}

\subsection{Recruited Collaborators Criteria Generation}
\label{Team Member Criteria Generation}

\begin{tcolorbox}[breakable]
\bfseries
\small
As a $<<<$\{identity\}$>>>$, you are asked to judge the relevance of a passage in relation to the given query, list the criteria for judging relevance between the query and a passage, assign weights to each criterion. 
\\\\\\
Given query: $<<<$\{query\}$>>>$
\\\\\\
Please give your output in JSON format with keys \textbackslash``Criteria\textbackslash'' and \textbackslash``Reason\textbackslash''.
Under the content of \textbackslash``Criteria\textbackslash'', please output the criteria and some explanation to it, your criteria should be correct, clear, executable, and easy to understand, the criteria should come from your knowledge as the $<<<$\{identity\}$>>>$. The weight to every criterion should be added after each criterion with 'The weight to this criterion is:'.
\\\\\\
Under the content of \textbackslash``Reason\textbackslash'', explain why you output these criteria.
\\\\\\
Focus solely on your identity and avoid deductions outside of your $<<<$\{identity\}$>>>$. 
\\\\\\
Provide a clear and concise response, just give your answer in JSON format as I request, don't say any other words.

\end{tcolorbox}

\subsection{Team Member Score}
\label{Team Member Score}
\begin{tcolorbox}[breakable]
\bfseries
\small
Ignore all previous instructions.
\\\\\\
Role: You are a $<<<$\{identity\}$>>>$
\\\\\\
Task Description: I will give you the criteria, one passage, and a query, you should follow the criteria to analyze the given passage's relevance from the view as a $<<<$\{identity\}$>>>$. Generate your answer in JSON format, with the key is \textbackslash``Score\textbackslash''. Under the content of \textbackslash``Score\textbackslash'', give me an integer score from 0 to 10 to represent the $<<<$\{identity\}$>>>$'s view relevance degree, just give the score, don't say any other words.
\\\\\\
Input-Output Description: The input will be the criteria, a passage, and a query, generate your output as I request:
\\\\\\
Your Answer:
\\\\\\
query: $<<<$\{query\}$>>>$
\\\\\\
criteria: $<<<$\{criteria\}$>>>$
\\\\\\
passage: $<<<$\{passage\}$>>>$
\\\\\\
Output:
\end{tcolorbox}

\subsection{LLM Assessor}
\label{LLM assessor}
\begin{tcolorbox}[breakable]
\bfseries
\small
Ignore all previous instructions.
\\\\\\     
Role: You are adept at synthesizing diverse viewpoints to reach a well-considered conclusion.
\\\\\\     
Task Desciption: Your task is to evaluate the relevance of a given passage to a specified query. You will receive a passage, a query, and some relevance assessments from experts. These experts come from different fields and may not always agree. After reviewing their assessments, you are to integrate their insights and determine a final relevance score. Please give your output in JSON format,with the key is \textbackslash``Final score\textbackslash''. Under the content of \textbackslash``Final score\textbackslash'' after carefully thinking about the two experts' scores, combine them and just give one final relevance score, the score should be an integer from 0 to 10, don't say any other words. Give your output in JSON as I request, don't explain,don't say any other words.
\\\\\\        
Input-Output Description: The input will be a passage, a query and three relevance assessments from three experts, generate your output as I request:
\\\\\\        
Your Answer:
\\\\\\      
query: $<<<$\{query\}$>>>$
\\\\\\        
passage: $<<<$\{passage\}$>>>$
\\\\\\        
Assessment of the three experts: 
Assessments from expert 1 who is an expert in  $<<<$\{identity\}$>>>$: 
Relevance Score 1 (From 0 to 10): $<<<$\{score\}$>>>$ 
\\\\\\
Assessments from expert 2 who is an expert in $<<<$\{identity\}$>>>$ : 
Relevance Score 2 (From 0 to 10): $<<<$\{score\}$>>>$
\\\\\\
Assessments from expert 3 who is an expert in language: 
Relevance Score 3 (From 0 to 10): $<<<$\{score\}$>>>$
\\\\\\        
Output:      
\end{tcolorbox}

\subsection{Rating Scale 0-to-k Directly Score (DIRECT(0, k))}
\label{DIRECT(0, k)}
\begin{tcolorbox}[breakable]
\bfseries
\small
Ignore all previous instructions.
\\\\\\     
Role: You are an expert in judging relevance.
\\\\\\     
Task Description: From a scale of 0 to 10, judge the relevance between the query and the document. Give your output in JSON format, with the key is \textbackslash``Score\textbackslash''. Under the content of \textbackslash``Score\textbackslash'', just give me one final integer score from 0 to 10, don't say any other words.
\\\\\\     
Input-Output Description: The input will be a query and a document.
\\\\\\     
Your Answer:
\\\\\\     
query: $<<<$\{query\}$>>>$
\\\\\\     
passage: $<<<$\{passage\}$>>>$
\\\\\\     
Output:

\end{tcolorbox}


\end{document}